\documentclass[
]{ceurart}

\usepackage{listings}
\usepackage{xcolor}

\begin{document}

\copyrightyear{2024}
\copyrightclause{Copyright for this paper by its authors.
  Use permitted under Creative Commons License Attribution 4.0
  International (CC BY 4.0).}

\conference{RecSys Workshop on Recommenders in Tourism (RecTour 2024), October 14th-18th, 2024, co-located with the 18th ACM Conference on Recommender Systems, Bari, Italy}

\title{TRACE: Transformer-based user Representations from Attributed Clickstream Event sequences}

\author[1]{William Black}[%
email=wblack@expediagroup.com,
]
\author[1]{Alexander Manlove}[%
email=amanlove@expediagroup.com,
]
\author[1]{Jack Pennington}[%
email=jpennington@expediagroup.com,
]
\author[1]{Andrea Marchini}[%
email=amarchini@expediagroup.com,
]
\author[1]{Ercument Ilhan}[%
email=eilhan@expediagroup.com,
]
\author[1]{Vilda Markeviciute}[%
email=vmarkeviciute@expediagroup.com,
]
\address[1]{Expedia Group, 407 St John St, London EC1V 4EX}

\begin{abstract}
  For users navigating travel e-commerce websites, the process of researching products and making a purchase often results in intricate browsing patterns that span numerous sessions over an extended period of time.
  The resulting clickstream data chronicle these user journeys and present valuable opportunities to derive insights that can significantly enhance personalized recommendations.
  We introduce TRACE, a novel transformer-based approach tailored to generate rich user embeddings from live multi-session clickstreams for real-time recommendation applications. 
  Prior works largely focus on single-session product sequences, whereas TRACE leverages site-wide page view sequences spanning multiple user sessions to model long-term engagement.
  Employing a multi-task learning framework, TRACE captures comprehensive user preferences and intents distilled into low-dimensional representations. 
  We demonstrate TRACE's superior performance over vanilla transformer and LLM-style architectures through extensive experiments on a large-scale travel e-commerce dataset of real user journeys, where the challenges of long page-histories and sparse targets are particularly prevalent.
  Visualizations of the learned embeddings reveal meaningful clusters corresponding to latent user states and behaviors, highlighting TRACE's potential to enhance recommendation systems by capturing nuanced user interactions and preferences.
\end{abstract}


\maketitle

\section{Introduction} 
\label{section:intro}

On tourism e-commerce websites users often exhibit complex navigation patterns whilst they browse travel and accommodation options before making a purchase. 
A typical user could land on the homepage, search for a flight then bounce, only to return a few days later to browse hotels and then purchase a package holiday.
The resulting clickstream data captures these intricate journeys and offers valuable insights into users' behaviour and intentions. 
By harnessing this data and better understanding users' latent psychological states and preferences, we can significantly enhance personalized experiences by matching them with more relevant content \cite{zhao2017recommending, shen2023learning, bcom_perso, Black_2023, grbovic2018real} and adapting the experience to better suit their context \cite{Black_2023}.
For instance, users earlier in their search can be presented with more exploratory content, as compared to users nearer the end of the purchase funnel.

However, achieving this level of personalization can be challenging as user journeys often span multiple sessions over an extended period of time, and specific goals, such as completing a purchase, occur infrequently within this window.
This is a particularly pertinent challenge within the tourism industry as users will often only make one booking a year, which can takes weeks of searching and planning before purchasing it months in advance.

In this work, we present TRACE (Transformer-based Representations of Attributed Clickstream Event sequences), a novel approach for generating rich user embeddings from live multi-session clickstream data with sparse targets.
TRACE employs a multi-task learning (MTL) framework, where a lightweight transformer encoder is trained to predict multiple user engagement targets based on sequences of attributed clickstream events.
By jointly predicting a diverse set of user future engagement signals, the model is encouraged to learn robust versatile representations.
We demonstrate its effectiveness using a real-world travel e-commerce dataset.

Numerous works have explored the use of statistical and machine learning techniques on clickstreams to mine patterns \cite{olmezogullari2020representation, bernhard2016clickstream, kim2011recommender} or cluster user behaviors \cite{wang2016unsupervised, su2015method, wei2012visual, zavali2021shopping} for analytical insights or motivating recommendations. Comparable works have also investigated neural and MTL approaches to user modeling, but typically focus on product-level interactions or single session sequences \cite{bai2023expressive, alves2022will, requena2020shopper}. TRACE instead ingests live clickstream data and addresses more general sequences of site-wide page views spanning multiple sessions in order to obtain rich user journey representations for real-time downstream applications.
PinnerFormer \cite{pancha2022pinnerformer} notably uses a transformer, but relies on previously learned embeddings and abundant pin-based interactions. TRACE learns directly and exclusively from the sequence of attributed page views and employs a MTL approach to overcome sparse engagement signals.
Zhuang et al. \cite{zhuang2019attributed} studied attributes at the sequence level, whereas TRACE is more granular and addresses attributes at the event-level. 
Where Rahmani et al. \cite{rahmani2023incorporating} incorporated temporal signals in sequential recommendations, TRACE instead adopts learnable positional encodings which capture both event and session positions. 

Overall, the key distinction of TRACE is the use of a transformer-based MTL framework with event-session position encoding to generate versatile user embeddings from enriched multi-session clickstream sequences with event-level attributes, which has not been explored in depth by previous research nor applied to travel e-commerce.

\section{Methodology}

\subsection{Problem Formulation}

Each time a user visits a new page it is logged in a clickstream as a page view event $P \in \mathcal{P}$, characterized by a small set of contextual features including the page name and timestamp $t_{P}$.
These events collectively form user sessions $\mathcal{S}$, representing ordered sequences of the pages visited within defined time intervals.
Formally, a session $\mathcal{S} = \{P_{0}, P_{1}, ..., P_{N}\}$, where $P_{j}$ denotes the $j$th page the user visited in this session, subject to the condition that
\begin{equation}
    t_{P_j} - t_{P_{j-1}} \leq T, \quad \forall j \in [1, N].
\end{equation}

Here $T$ is a fixed constant, often in the order of magnitude of a few hours. 
If the difference in timestamps of two sequential page view events is greater than $T$, the latter is considered to be in a new session.

Then for each user, we define their journey $J$ as the chronological sequence of their sessions, where $J = \{\mathcal{S}_0, \mathcal{S}_1, ..., \mathcal{S}_k\}$, with $\mathcal{S}_i$ representing their $i$th session. 
In this way a journey $J$ is the sequence of pages a user has visited across multiple sessions. 
We use a corpus of user journeys $\mathcal{J} = \{J_0, J_1,...\}$ captured on a large-scale travel e-commerce site over a few months, where $|\mathcal{J}| > 50$M, and the vocabulary exceeds $1000$ page names. 

Our objective is to predict future engagement of users using their past navigation patterns on the website.
Formally, we want to learn a model $ f:\mathcal{J} \rightarrow \mathbb{R}^{d}$ for some positive integer $d$, which summarises these journeys in rich low dimensional representations that can then be used for downstream machine learning applications, such as content personalisation and product recommendations. 
As such the model $f$ must satisfy three main requirements:

\begin{enumerate}
    \item Effectively capture the intricate page navigation patterns in users' journeys which span multiple sessions.
    \item Meaningfully distill user journeys into embeddings that can predict engagement across diverse tasks and contexts.
    \item Scale efficiently to accommodate high-traffic real-time production environments.
\end{enumerate}

To generate our datasets, we split each journey at a random point and designate the pages before as the input journey, and those after to be used for target generation. 

\begin{figure}[t]
  \centering
  \includegraphics[width=\linewidth]{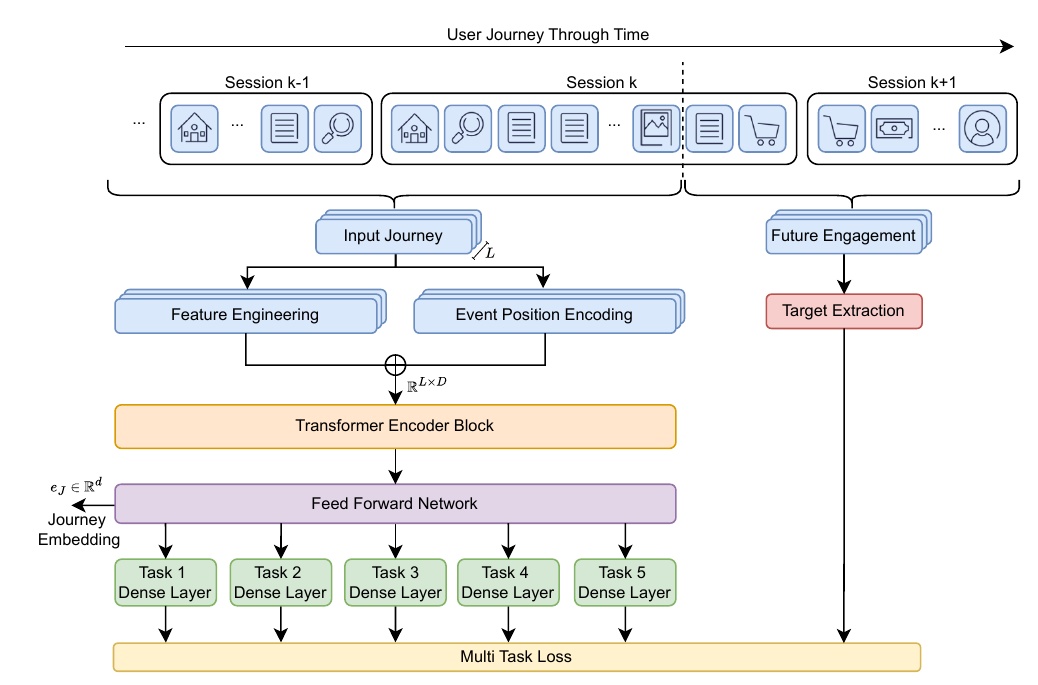}
  \caption{Overview of the TRACE multi-task transformer architecture.}
  \label{fig:mtl_architecture}
\end{figure}

In our proposed approach, TRACE, we train a multi-task transformer. This model takes as input some sequence of pages in the form of some journey $J$, and predicts a cohort of future user engagement targets. 
We extract the output of the final layer of the shared backbone of the model as the journey embedding $e_{J} \in \mathbb{R}^d$.
We hypothesise that if the embeddings are predictive across a cohort of diverse user engagement tasks, they will capture a generalised understanding of a user's diverse intents. Figure \ref{fig:mtl_architecture} illustrates the components of TRACE. We address each in more detail below.

\subsection{Feature and Position Encoding} \label{section:pos_enc}

We first crop each input journey, taking up to the $L$ most recent page view events, where $L$ is chosen in a way to capture most users' entire recent page view history. 

Each page view event $P$ has a set of categorical attributes, such as a page name and the user's device type, which are passed through their own learnable embedding layer to produce a dense representation in $\mathbb{R}^{32}$. 
We engineer two features from the event timestamp; the time interval between consecutive events and the time elapsed until the most recent event, both logged and standard scaled.
Additionally we encode session ID where events in the $n$th most recent session are given value $n$. 
These time-based features aim to capture planning phases and session gaps common in extended travel user journeys.
All features are standard scaled and concatenated s.t. each $P$ is now represented by a vector in $\mathbb{R}^{D}$, where $D$ is approximately a few hundred.

We also enumerate the event position, where the $m$th most recent event in the entire journey is given value $m$. 
Then both the event and session position indexes are independently embedded in $\mathbb{R}^{D}$ via their own learnable layers, and added onto the final feature vector, acting as an event-session position encoding. 
This was designed to allow the model to learn representations specific to session and position combinations, enabling it to capture dynamics both within and across multiple sessions more effectively.

For input journeys of length less than $L$ we pad the $D$ features with value 0. As such each journey $J$ can now be encoded as some matrix $M_{J} \in \mathbb{R}^{L \times D}$.

\subsection{Model Architecture}
 
In TRACE, we use a transformer encoder architecture to process the input sequences of pages, and train it in a multi-task regime across five different targets, representing a variety of future user engagement signals. 

An encoded journey $M_{J} \in \mathbb{R}^{L \times D}$ is passed through a single transformer encoder block, constisting of a multi-head self-attention layer with 8 heads followed by a position-wise fully connected feed-forward network (FFN) with an intermediate dimension of 128. 
We employ dropout and a residual connection around each of the two sub-layers, followed by layer normalization. 
Global max pooling is applied to the output of this encoder block, before being forward passed through a FFN.
For an input journey $J$ the output of this shared backbone is some $e_{J} \in \mathbb{R}^d$.

This tensor $e_{J}$ is then passed through five separate task-specific dense layers, each compressing down to a scalar value so the final output of the model is some five logits $\hat{\textbf{y}} \in \mathbb{R}^5$.
After training we then remove the five task-specific heads, and take the output of the shared backbone $e_{J}$ as the journey embedding.
We deliberately restrict the heads to be simple logistic regression layers. 
This approach encourages the shared backbone to capture most of the nuance, ensuring the embeddings are information-rich and generalizable, as opposed to relying too heavily on the task-specific layers.

Throughout the architecture we use $\mathrm{ReLU}$ activations, except for the final shared dense layer where sigmoid is used for its desirable bounding property.
This ensures normalization of the output embedding, with our experiments demonstrating no performance loss.
We set dimension $d = 32$ for the embedding, which is well suited for downstream applications.

\subsection{Multi Task Training Regime and Objective}

The motivation behind the MTL approach is that by jointly predicting a diverse set of user engagement signals, the model is encouraged to learn comprehensive and generalizable representations that can be effectively utilized across a variety of downstream applications, extending beyond just the tasks during training.
Furthermore, by mixing the infrequent targets such as purchases, with more common events like product searches, the model learns from a stronger signal and as our results show perform better on those sparse tasks.
This is especially advantageous in the travel domain for events such as bookings, as demonstrated in our experiments.

The model is trained on five binary classification tasks which represent potential future actions of a user: (PW2) Make any purchase within two weeks; (BN5) Bounce within next five pages, and the following which relate to actions within rest of session; (SRP) Make a search for a product; (PDP) View a product details page; and (VUO) View an upcoming order.
Each task head has its own class-weighted binary cross-entropy loss function.
The overall objective is expressed as a linear combination of these task-specific losses.
For a journey $J$ with model prediction $\hat{\textbf{y}}$ and true labels $\textbf{y}$, the loss is defined as: 
\begin{equation} \label{eqn:loss_function}
    \mathcal{L}(J, \textbf{y}) = - \sum_{k=1}^5 \left[ w_k \cdot y_k \cdot \log(\hat{y}_k) + (1 - y_k) \cdot \log(1 - \hat{y}_k) \right].
\end{equation} 
Class weights $w_k$ are computed as the reciprocal of the proportion of positive samples for each task $k$, in order to account for task-specific class imbalance.
We weights tasks equally to encourage the model to develop features which generalize across each task.

\section{Experimental Results}

\subsection{Downstream Embedding Evaluation} \label{section:downstream_embedding_eval_strategy}

Supervised probing techniques have previously been developed to assess linguistic embeddings \cite{tenney2019you, hewitt2019structural, hewitt2019designing}, although are not directly suited to this scenario. We instead propose a downstream strategy for evaluating the richness of information contained within a set of embeddings. After training, we compute ground truth targets on an unseen test set of historical user journeys. These targets seek to encapsulate users' latent psychological states and future intentions. For this, we use the same five tasks from the TRACE objective in eqn. \ref{eqn:loss_function}, and introduce three more evaluation tasks that were not previously seen. These include: (HOM) whether a user returns to the homepage; (PWS) converts within the current session; and (RE7) whether they return to the site within seven days. This captures a broad scope of user outcomes, allowing us to characterize how well the embeddings generalize. 
We pass the unseen test journeys through the model to obtain a corresponding set of embeddings. 
Next, we train XGBoost models \cite{chen2016xgboost} on these test set embeddings. We fit one XGBoost model independently to each unique evaluation task and optimize hyperparameters, such as \textit{max\_depth} and \textit{learning\_rate}, using K-fold cross validation.
The trained XGBoosts then undergo evaluation and we compute performance metrics on the model predictions.
These metrics serve as proxies for assessing the richness of embeddings and exemplify downstream model performance across various use cases. Throughout this section, we evaluate each upstream embedding model by using the same procedure on the same unseen test set.

\subsection{Comparable Models vs TRACE}

We evaluate the quality of TRACE embeddings against several comparable approaches. We express our comparisons as the mean uplift taken over all evaluation tasks. Results are shown in Table \ref{tab:model_vs_baselines}.

\textbf{Myopic Baseline.} Our baseline predicts targets using explicit attributes from only the most recent event. 
We report all results as percentage uplifts from this.
TRACE significantly outperforms this, highlighting the benefits of mining a user's full navigation history.

\textbf{Single Task Cohort.} To demonstrate the effectiveness of TRACE's MTL approach, we trained a dedicated single-task transformer for each of the evaluation tasks. These models each produce an embedding.
In Table \ref{tab:trace_vs_singletask}, the TRACE score on a given task is compared to the corresponding dedicated ST model embedding's score.
Overall results show that the TRACE embeddings outperform every task-specific equivalent on the 5 tasks TRACE was trained on, and even wins on all but one of the unseen targets, demonstrating the advantages of the MTL approach.

\textbf{Single Task Aggregated.} Here we combine the task-specific models' embeddings into a single embedding of the same length by taking the mean along each dimension.

\textbf{Multi-Task LSTM.} We note the demonstrated efficacy of LSTMs in related works \cite{alves2022will, zhuang2019attributed, koehn2020predicting, sakar2019real}.
We train a comparable LSTM minimizing the same multi-task objective function shown in (\ref{eqn:loss_function}).

\textbf{Mini-GPT.} We train a small GPT-style model \cite{radford2018improving} on the page name sequences, with a single transformer block and causal masking in the attention layer for next event prediction. Embeddings are computed from the mean of the transformer block outputs.

\begin{table}[h]
    \renewcommand{\arraystretch}{0.95}
    \centering
    \caption{Mean \% uplift in XGBoost metrics from myopic baseline across eval. tasks for TRACE and comparable models.}
    \begin{tabular}{l | r r r r}
    \toprule
        \textbf{Model} & \textbf{AUROC} & \textbf{AUPRC} & \textbf{F1} & \textbf{Acc} \\
    \midrule
        \textbf{TRACE} & \textbf{+7.23} & \textbf{+13.58} & \textbf{+2.73} & \textbf{+2.15} \\
        ST Cohort & $+6.38$  & $+10.75$  & $+2.72$  & $+2.06$  \\
        ST Aggregated & $+6.34$  & $+10.62$  & $+2.18$  & $+1.73 $ \\
        MT LSTM & $+1.91$  & $-3.29$  & $-0.29$  & $+0.27$  \\
        Mini-GPT & $+1.86$  & $-2.40$  & $-1.13$  & $-0.60$  \\
    \bottomrule
    \end{tabular}\\[2pt]
    \label{tab:model_vs_baselines}
\end{table}

\begin{table}[h]
    \renewcommand{\arraystretch}{1.0}
    \centering
    \caption{Mean \% uplift in XGBoost AUROC on the eight eval. tasks for TRACE vs dedicated single-task models.}
    \begin{tabular}{p{.115\linewidth} | p{.064\linewidth} p{.064\linewidth}  p{.064\linewidth} p{.064\linewidth} p{.064\linewidth} | p{.064\linewidth}
    p{.064\linewidth} p{.064\linewidth}}
    \toprule
        \textbf{Model} & \textbf{PW2} & \textbf{BN5} & \textbf{SRP} & \textbf{PDP} &  \textbf{VUO} & \textbf{HOM} & \textbf{PWS} & \textbf{RE7} \\
    \midrule
        \textbf{TRACE} & \textbf{+11.8} & \textbf{+9.32} & \textbf{+6.73} & \textbf{+7.29} & \textbf{+3.99} & \textbf{+7.70} & \textbf{+4.35} & +6.52\\
        ST & +11.2 & +8.14 & +5.08 & +6.25 & +3.56 & +5.59 & +2.99 & \textbf{+8.26} \\
    \bottomrule
    \end{tabular}\\[2pt]
    \label{tab:trace_vs_singletask}
\end{table}

\subsection{Visualisation of Learned Embeddings}

\begin{figure}[h]
  \centering
  \includegraphics[width=0.7\linewidth]{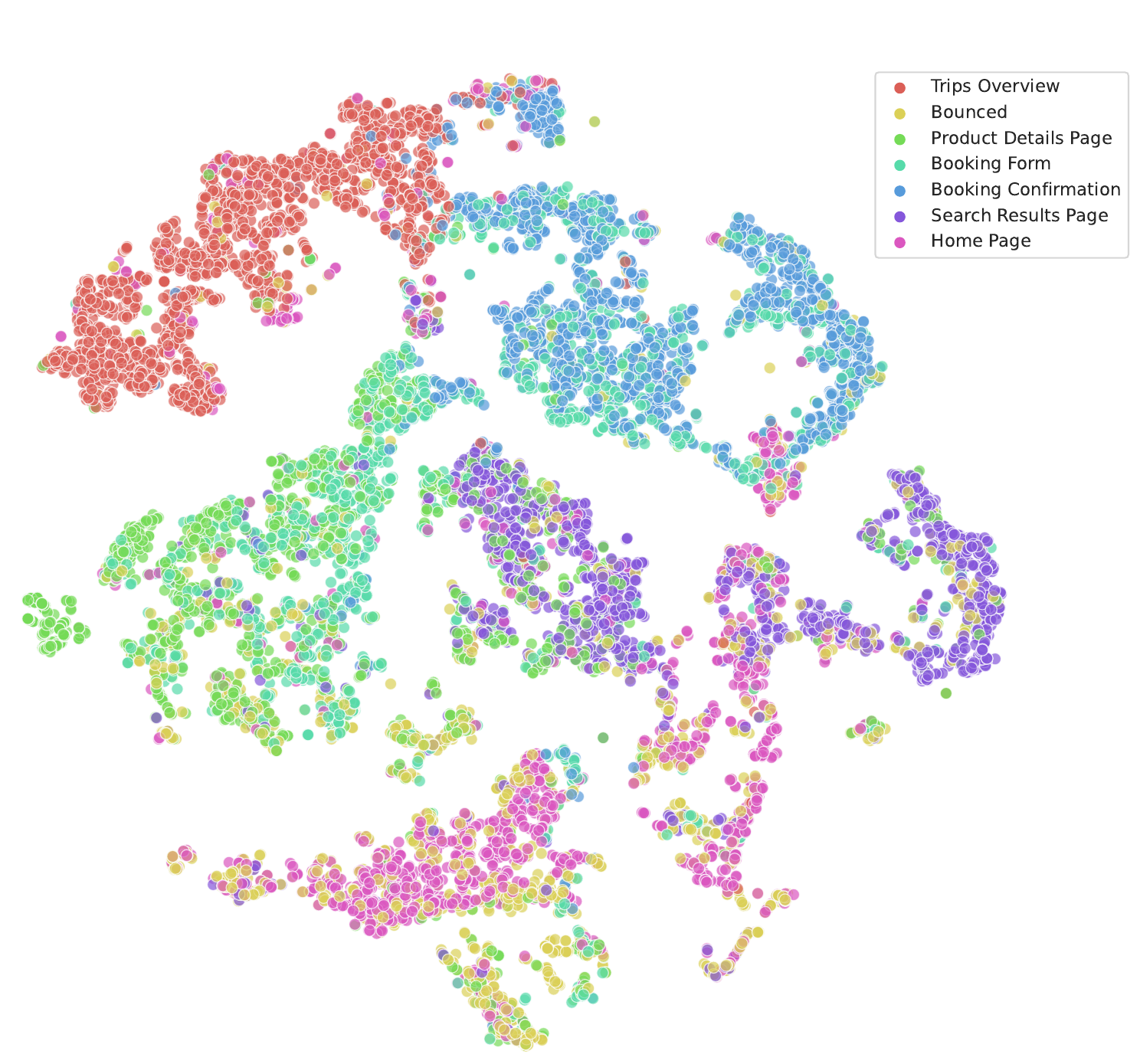}\\[1pt]
  \caption{t-SNE projections of TRACE page sequence embeddings, colored according to the next page the users visited.}
  \label{fig:tsne_scatterplot}
\end{figure}

In Fig. \ref{fig:tsne_scatterplot}, we present a visualization of the 32-dimensional embeddings learned by TRACE, reduced to 2 dimensions using t-SNE \cite{van2008visualizing}.
This subset of observations was uniformly sampled with respect to users' next visited page, ensuring equal representation from seven common pages.
We note the emergence of clusters corresponding to the next page visited by users, despite TRACE never being explicitly exposed to this information during training.
Qualitatively, the clusters appear to loosely align with how a user traverses a website, going from homepage at the bottom progressing through to search and product pages, before reaching checkout and order confirmation.  
This underscores TRACE's ability to identify and encode patterns in user journeys, showcasing the effectiveness of our approach for generating information-rich embeddings. 

\subsection{Ablation Experiments}

In Table \ref{tab:ablation_study_results} we list the results of our ablation studies.

\subsubsection{Position Encodings} \label{subsub:pos_enc}

In section \ref{section:pos_enc}, we discussed our approach to position encoding which is designed to tackle event sequences over multiple sessions. 
Static trigonometric position encodings are also widely popular \cite{bai2023expressive, vaswani2017attention}. 
We trained a variant including this additional encoding, but found better performance without it.

\subsubsection{Number of Encoders} \label{subsub:num_encs}

Here, we vary the number of transformer encoders, $h$. 
Our results suggest that $h=1$ encoder is sufficient for capturing the structure of the data, likely due to our sequences being of relatively shorter length with small vocabulary compared to typical NLP applications \cite{vaswani2017attention}.
We measured the time taken for the forward pass in each variant. 
The experiments were conducted on a system equipped with an Nvidia T4 Tensor Core GPU (16 GiB VRAM) and an Intel Xeon processor (32 vCPUs, 128 GiB RAM, 2.5 GHz clock speed). 
We repeat the model call 10,000 times and measure the mean and standard deviation for various encoder configurations. The results are as follows:

\begin{itemize}
    \item h = 1 encoder: 27.5 ms ± 0.1 ms
    \item h = 2 encoders: 40.8 ms ± 0.1 ms
    \item h = 3 encoders: 54.7 ms ± 0.6 ms
    \item h = 4 encoders: 67.7 ms ± 0.4 ms
\end{itemize}
 
Our final model design used only a single encoder h=1, which is sufficiently fast taking only 27.5 milliseconds on average for the forward pass. 
This is well within our self-imposed upper limit of 100ms latency, which we find to be practical for real-time applications.

\subsubsection{Chronological Features} \label{subsub:session_id_time_feat}
To better understand the specific performance gains from chronological features, we train variants which omit these. 
The "No Session ID" variant retains event timestamps but removes session ID, thereby eliminating explicit information about session continuity.
The "No Time" variant excludes both session IDs and timestamps, retaining only the sequential order of events. 
Results demonstrate that including timestamp features enhances performance, but the greatest improvement arises from incorporating TRACE's session encoding on top of this, as used in our final variant.
This highlights the effectiveness of TRACE in exploiting the multi-session structure of the sequences, and its significance for applications in e-commerce recommendation systems.

\begin{table}[t]
    \renewcommand{\arraystretch}{.91}
    \centering
    \caption{Mean XGBoost performance uplift \% on evaluation tasks for TRACE variants' embeddings vs myopic baseline.}
    
    \begin{tabular}{p{.1\linewidth} p{.035\linewidth} p{.14\linewidth} | p{.11\linewidth} p{.11\linewidth} p{.11\linewidth} p{.1\linewidth}}
    \toprule
        \textbf{Pos. Enc.} & \textbf{$h$} & \textbf{Time Feats.} & \textbf{AUROC} & \textbf{AUPRC} & \textbf{F1} & \textbf{Acc.}\\
    \midrule
        Event$^\dagger$ & 1$^\dagger$ & All$^\dagger$ & \textbf{+7.23 } & \textbf{+13.58 } & \textbf{+2.73 } & $+2.15$   \\
    \midrule
        Trig. &  &  & $+6.64$  & $+12.22$  & $+2.47$  & \textbf{+2.17 }  \\
    \midrule
         & 2 &  & $+6.87$  & $+12.62$  & $+2.65$  & $+2.07$  \\
         & 3 &  & $+6.84$  & $+12.76$  & $+2.53$  & $+1.98$   \\
         & 4 &  & $+6.84$  & $+12.76$  & $+2.61$  & $+2.06$  \\
    \midrule
         &  & No Sess. & $+6.62$  & $+12.24$ & $+2.45$  & $+1.94$   \\
         &  & No Time & $+6.36$ & $+12.0$ & $+2.17$  & $+1.70$   \\
    \bottomrule
    \end{tabular}
    \label{tab:ablation_study_results}\\
    $^\dagger$\textit{Final variant used in proposed TRACE model}.
\end{table} 

\section{Conclusion}

In this work we have presented TRACE, a novel approach for generating user embeddings from multi-session page view sequences through a multi-task learning (MTL) framework, which employs a lightweight encoder-only transformer to process real-time cross-session clickstream data.
Our experiments on a large scale real world travel e-commerce dataset, demonstrate the superior performance of TRACE embeddings compared to traditional single-task and LSTM-based models, and highlights its potential for enhancing tourism recommender systems.
The learned embeddings exhibit strong results on a diverse set of targets and demonstrate the ability to generalize well to unseen tasks, underscoring their utility for applications like content personalization and user modeling. 
Visualizations reveal that TRACE can effectively capture meaningful clusters corresponding to latent user intents and behaviours.

In future, we plan to explore the integration of LLMs, as in \cite{christakopoulou2023large, fan2023llmrecommender}, and investigate hierarchical models to further improve the model's representational capacity.

\newpage

\bibliography{paper_bibliography}

\end{document}